\documentclass[]{cupconf}
\usepackage{graphicx}

\title[[Formation \& evolution of the Galactic bulge]
      {Formation and evolution of the Galactic bulge: \\ constraints
      from stellar abundances} 
      
\author[S.K. Ballero et al.]{\\ \ 
\\ Silvia K. Ballero$^{1,2}$ 
\\ Francesca Matteucci$^{1,2}$
\\ Livia Origlia$^3$}
\affiliation{$^1$ Dipartimento di Astronomia, Universit\`a di Trieste,
via G.B. Tiepolo 11, 34124 Trieste, Italy\\ 
$^2$ INAF-Osservatorio Astronomico di Trieste, Via G.B. Tiepolo 11, I-34121 
Trieste, Italy \\ 
$^3$ INAF-Osservatorio Astronomico di Bologna, Via G. Ranzani 1, I-40127 
Bologna, Italy}

\begin{document}
\maketitle

\begin{abstract}
We present results for the chemical evolution of the Galactic bulge in
the context of an inside-out formation model of the Galaxy. 
A supernova-driven wind was also included in analogy with elliptical
galaxies. 
New observations of chemical abundance ratios and
metallicity distribution have been employed in order to check the
model results. 
We confirm previous findings that the bulge formed on a
very short timescale with a quite high star formation efficiency and
an initial mass function more skewed toward high masses than
the one suitable for the solar neighbourhood. 
A certain amount of primary nitrogen from massive stars might
be required to reproduce the nitrogen data at low and intermediate
metallicities.
\end{abstract}

\firstsection
\section{Introduction}
The issue of bulge formation and evolution has lately received
renewed attention due to recent studies which form the basis for a
growing consensus that the bulge is old and that its formation
timescale was relatively short ($\leq 1$ Gyr).
We want to test the hypothesis of a quick dissipational collapse via
the study of the evolution of the abundance ratios coupled with
considerations on the metallicity distribution and to show that
abundance ratios can provide an independent constraint for the bulge
formation scenario since they differ depending on the star formation
history (Matteucci, 2000). 
$\alpha$-elements in particular are of paramount importance in probing
the star formation timescale, since the signature of a very short
burst of star formation must result into an enhancement with respect
to iron.

\section{The chemical evolution model and the data}
The model we are adopting is described by Ballero et
al. (2006b) and belongs to the category of fast dissipational collapse
prior to the settling of the disk. It follows the trend of previous
models by Matteucci \& Brocato (1990) and Matteucci et al. (1999). The
star formation rate is proportional to the gas surface mass density of
the disk:
\begin{equation}
\psi(r,t) = \nu G(r,t)
\end{equation}
where $\nu$ is the star formation efficiency, i.e. the inverse of the
star formation timescale. The stellar initial mass function (IMF) is
parametrized as a power-law of logarithmic index $x$:
\begin{equation}
\phi(m) \propto m^{-(1+x)}
\end{equation}
which may differ among the various mass ranges.
Finally, the bulge forms via infall of gas shed from the Galactic halo
on a timescale $\tau$:
\begin{equation}
\dot{G}_{inf}(t) \propto e^{-t/\tau}
\end{equation}
We updated the stellar lifetimes by adopting those of Kodama (1997)
and we introduced the stellar yields by Fran\c cois et al. (2004)
which were calibrated in order to fit the chemical properties of the
solar neighbourhood. Following the photometric measurements of Zoccali
et al. (2000) the IMF index below $1 M_{\odot}$ was set $x=0.33$. 
The Type Ia supernova rate is computed following the single-degenerate 
scenario, according to Matteucci \& Recchi (2001). Finally, a primary
production of nitrogen from massive stars, as computed by Matteucci
(1986) is considered as opposed to purely secondary production, which
is usually adopted in chemical evolution model.
We also introduced a supernova-driven wind in analogy with elliptical
galaxies. To develop it, we supposed that the bulge lies at the bottom
of the potential well of the Galactic halo, whose mass is $M_{\odot} =
10^{12}M_{\odot}$ and whose effective radius is $\sim100$ times the
effective radius of the bulge mass distribution. 
The binding energy of the bulge is then calculated following Bertin et
al. (1992), who split the potential energy into two parts, one due to
the bulge potential to bulge gas interaction and the other due to the
interaction of the bulge gas with the potential well of the dark
matter halo of the Milky Way, which depends on the relative mass
distribution.
The thermal energy of the interstellar medium due to supernova
explosions is given by:
\begin{equation}
E_{th,SNI/II}=\int_0^t\epsilon(t-t')R_{SNI/II}(t')dt'
\end{equation}
where $R$ is the rate of Type Ia and Type II supernovae, $t'$ is the
explosion time and $\epsilon$ is the energy content of a supernova
remnant, whose evolution is calculated following Cox (1972) and Cioffi
\& Shull (1988). When $E_{th,SNI/II}=E_{b,gas}$ a wind develops which
devoids the bulge of all its gas, and the evolution is passive
thereafter.

\begin{figure}
\centering
\includegraphics[width=.45\textwidth]{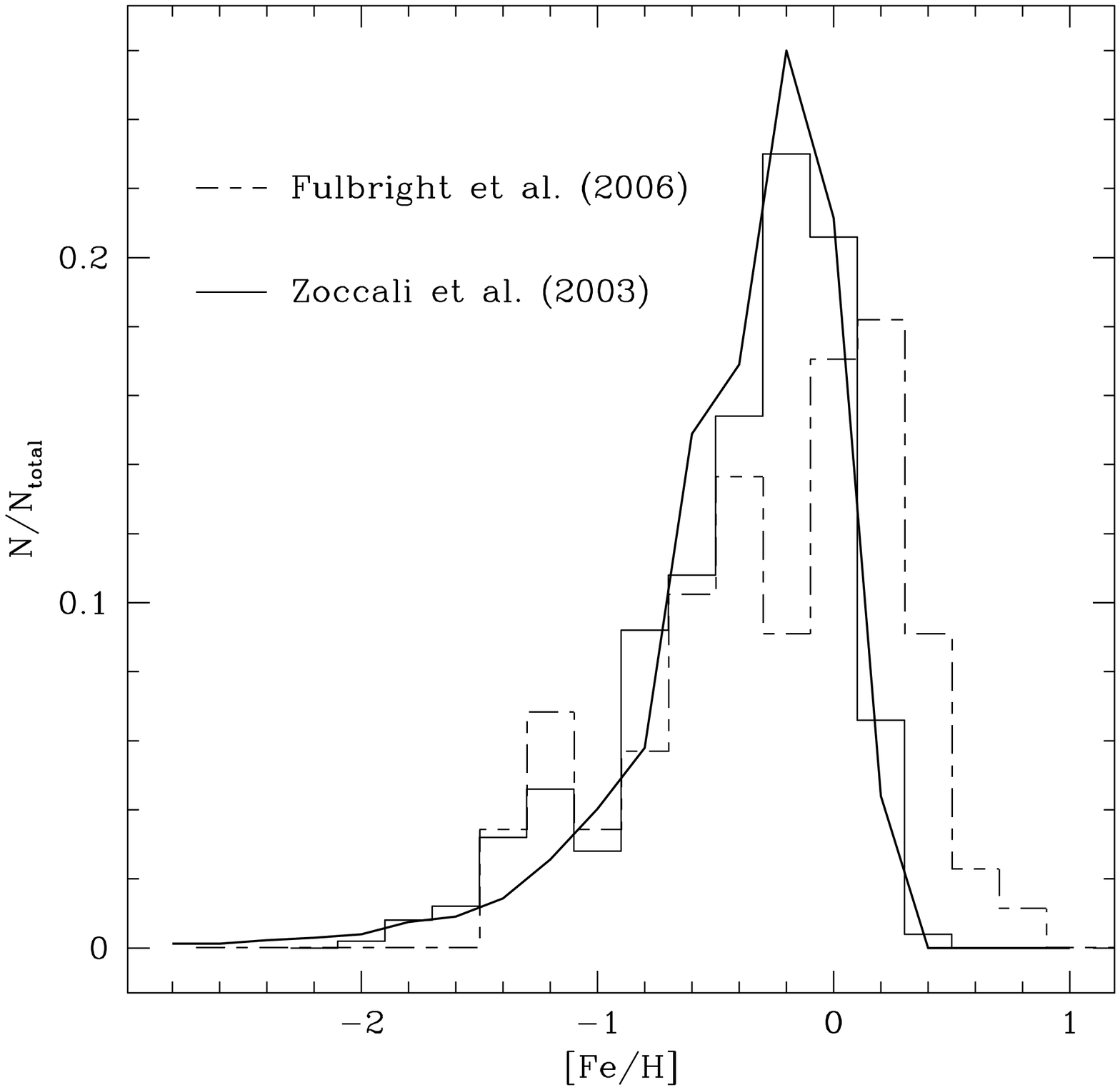}%
\includegraphics[width=.45\textwidth]{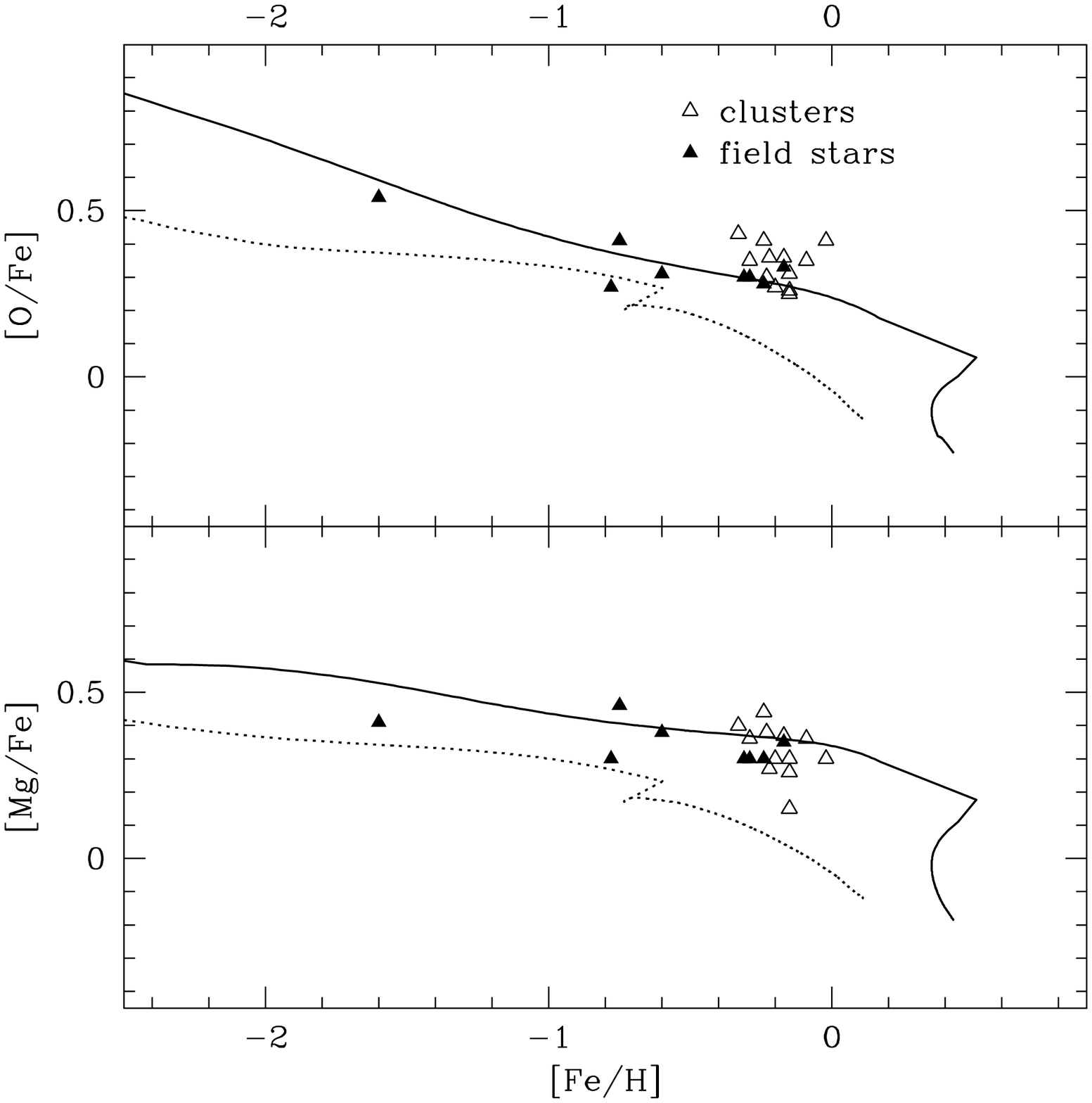}
\caption{\emph{Left panel}:
  Predicted metallicity distribution in our bulge reference model
  ($x=0.95$ for $M > 1M_{\odot}$, $\nu=20$ Gyr$^{-1}$, $\tau=0.1$ Gyr)
  compared with the observed distributions (Zoccali et al., 2003;
  Fulbright et al., 2006).
  \emph{Right panel}:
  Evolution of [O/Fe] and [Mg/Fe] vs. [Fe/H] in the bulge for
  the reference model (solid line). 
  A solar neighbourhood fiducial line (dotted line) is plotted for
  comparison. 
  Data are taken from the IR spectroscopic database (Origlia et al.,
  2002; Origlia \& Rich, 2004; Origlia et al., 2005; Rich \& Origlia,
  2005).}
\label{fig:fig1}
\end{figure}

\begin{figure}
\centering
\includegraphics[width=.85\textwidth,clip,trim=0 275 0 0]{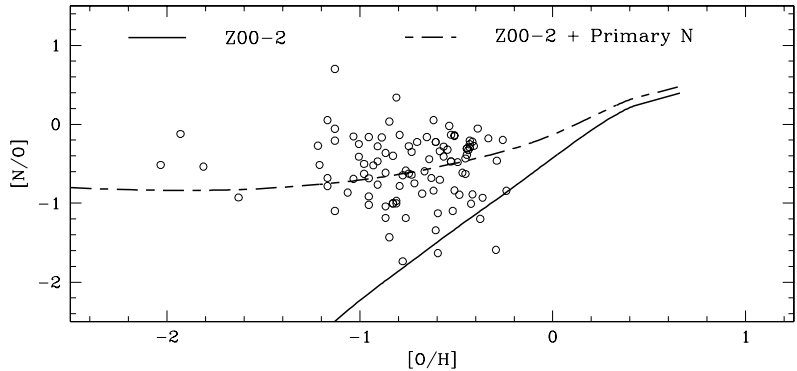}
\caption{Evolution of [N/O] vs. [O/H] in the Galactic bulge in our
  reference model (solid line) compared to a model where primary
  production of N from massive stars is assumed (Matteucci, 1986,
  dashed line).
  If the average trend of these observations is representative of the 
  pristine [N/O] values, the fit is achieved with the latter model. 
  Data for N and O in bulge PNe are from G\'orny et al. (2004).}
\label{fig:fig2}
\end{figure}

\section{Discussion}

We investigated variations of the parameters which influence the
chemical evolution, namely of the IMF index (from 0.33 to the Scalo
(1998) value) of the star formation efficiency $\nu$ (from $2$ to
$200$ Gyr$^{-1}$) and of the infall timescale $\tau$ (from $0.01$ to
$0.7$ Gyr) and matched the results with the most recent data for the
metallicity distribution and the evolution of [$\alpha$/Fe] abundance
ratios with metallicity (see Fig. \ref{fig:fig1}).
We found out that in order to reproduce all the constraints it is
necessary to adopt a short formation timescale ($0.01-0.1$ Gyr) and an
intense efficiency of star formation ($10-20$ Gyr), combined with an
IMF flatter ($x=0.95$ for $M>1M_{\odot}$) than that suitable for the
solar neighbourhood. 
The reference model we plotted is also able to explain the
different trends of [O/Fe] and [Mg/Fe] as seen from the observations. 
Moreover, if we compare (Fig. \ref{fig:fig2}) the [N/O] vs. [O/H] plot
with the data from planetary nebulae of G\'orny et al. (2004), we see
that in order to reproduce the average trend of observations it is
necessary to assume a primary production of nitrogen from massive
stars of all masses at every metallicity, in analogy with the solar
neighbourhood (Ballero et al, 2006a). 
The most promising way seems to be constituted by the rotational
yields of Meynet et al. (2006), as was shown in fact by Chiappini et
al. (2006).
In any case, conclusions regarding nitrogen cannot
be drawn firmly due to the fact that nitrogen is self-enriched to some
extent in planetary nebulae.


\begin{thebibliography}{99}

\bibitem[]{ballero:2006a} 
  Ballero, S.K., Matteucci, F., Chiappini, C. (2006a). 
  \textit{NewA} \textbf{11}, 306-324. 

\bibitem[]{ballero:2006b} 
  Ballero, S.K., Matteucci, F., Origlia, L., Rich, R.M. (2006b). 
  submitted to \textit{AJ}

\bibitem[]{bertin:1992} 
  Bertin, G., Saglia, R.P., Stiavelli, M. (1992). 
  \textit{ApJ} \textbf{384}, 423-432.

\bibitem[]{chiappin:2006} 
  Chiappini, C., Hirschi, R., Meynet, G., Ekstr\"om, S., Maeder, A.,
  Matteucci, F. (2006).
  \textit{A\&A} \textbf{449}, L27-L30. 

\bibitem[]{francois:2004} 
  Fran\c cois , P., Matteucci, F., Cayrel, R., Spite, M., Spite, F.,
  Chiappini, C. (2004).
  \textit{A\&A} \textbf{421}, 613-621.

\bibitem[]{fulbright:2006} 
  Fulbright, J.P., McWilliam, A., Rich, R.M. (2006).
  \textit{ApJ} \textbf{636}, 821-841.

\bibitem[]{gorny:2004} 
  G\'orny, S.K., Stasi\'nska, G., Escudero, A.V., Costa,
  R.D.D. (2004).
  \textit{A\&A} \textbf{427}, 231-244.

\bibitem[]{kodama:1997} 
  Kodama, T. (1997). 
  PhD thesis, University of Tokio.

\bibitem[]{meynet:2006} 
  Meynet, G., Ekstr\"om, S., Maeder, A. (2006).
  \textit{A\&A} \textbf{447}, 623-639.

\bibitem[]{matteucci:1986} 
  Matteucci, F. (1986). 
  \textit{MNRAS} \textbf{221}, 911-921.

\bibitem[]{matteucci:2000} 
  Matteucci, F. (2000).
  in ``The Evolution of the Milky Way: Stars versus Clusters'', 
  eds. F. Matteucci \& F. Giovannelli, 
  \textit{A.S.S.L.} vol. \textbf{255}, 3-12.

\bibitem[]{matteucci:1990} 
  Matteucci, F., Brocato, E. (1990).
  \textit{ApJ} \textbf{365}, 539-543.

\bibitem[]{matteucci:1999} 
  Matteucci, F., Romano, D., Molaro, P. (1999)
  \textit{A\&A} \textbf{341}, 458-468.

\bibitem[]{origlia:2004} 
  Origlia, L., Rich, R.M. (2004).
  \textit{AJ} \textbf{127}, 3422-3430.

\bibitem[]{origlia:2002} 
  Origlia, L., Rich, R.M., Castro, S. (2002).
  \textit{AJ} \textbf{123}, 1559-1569.

\bibitem[]{origlia:2005} 
  Origlia, L., Valenti, E., Rich, R.M. (2005).
  \textit{MNRAS} \textbf{256}, 1276-1282. 

\bibitem[]{rich:2005} 
  Rich, R.M., Origlia, L. (2005).
  \textit{ApJ} \textbf{634}, 1293-1299.

\bibitem[]{scalo:1998} 
  Scalo, J.M. (1998).
  in ``The Stellar Initial Mass Function'', 
  eds. G. Gilmore \& D. Howell, 
  \textit{ASP Conf. Series} \textbf{142}, 201-236.

\bibitem[]{zoccali:2000} 
  Zoccali, M., Cassisi, S., Frogel, J.A., Gould, A., Ortolani, S.,
  Renzini, A., Rich, R.M., Stephens, A.W. (2000).
  \textit{ApJ} \textbf{530}, 418-428.

\bibitem[]{zoccali:2003} 
  Zoccali, M., Renzini, A., Ortolani, S., Greggio, L., Saviane, I.,
  Cassisi, S., Rejkuba, M., Barbuy, B., Rich, R.M., Bica, E. (2003).
  \textit{A\&A} \textbf{399}, 931-956.  

\end{thebibliography}
\end{document}